\pdfoutput=1
\documentclass[conference]{IEEEtran}
\IEEEoverridecommandlockouts

\usepackage{booktabs} 
\usepackage{comment}
\usepackage{amsmath}
\usepackage{amssymb}
\usepackage{breakurl}
\usepackage{subfigure}
\usepackage{pifont}
\usepackage{graphicx}
\usepackage{color}
\usepackage{ifthen}
\usepackage{multirow}
\usepackage{url}
\usepackage{balance}

\newcommand{\sect}[1]{\textsection#1}

\DeclareMathOperator*{\argmin}{arg\,min}

\renewcommand{\figurename}{Fig.}

\def\BibTeX{{\rm B\kern-.05em{\sc i\kern-.025em b}\kern-.08em
    T\kern-.1667em\lower.7ex\hbox{E}\kern-.125emX}}

\usepackage{ifluatex}
\ifluatex
\usepackage{hyperref}
\usepackage{spelling}
\spellingreadgood{/misc/tools/dictionaries/whitelist}
\spellingreadgood{/misc/tools/dictionaries/whitelist2}
\fi

\newcommand{\inlineeqnum}{\refstepcounter{equation}~~\mbox{(\theequation)}}

\begin{document}

\title{Attack-Aware Data Timestamping in Low-Power Synchronization-Free LoRaWAN}

\author{\IEEEauthorblockN{Chaojie Gu$^\ast$ $\qquad$ Linshan Jiang$^\ast$ $\qquad$ Rui Tan$^\ast$ $\qquad$ Mo Li$^\ast$ $\qquad$ $\qquad$ Jun Huang$^\dag$}
\IEEEauthorblockA{$^\ast$\textit{Nanyang Technological University, Singapore} $\qquad$ $^\dag$\textit{Peking University, China}}}

\maketitle

\begin{abstract}
  Low-power wide-area network technologies such as LoRaWAN are promising for collecting low-rate monitoring data from geographically distributed sensors, in which timestamping the sensor data is a critical system function. This paper considers a synchronization-free approach to timestamping LoRaWAN uplink data based on signal arrival time at the gateway, which well matches LoRaWAN's one-hop star topology and releases bandwidth from transmitting timestamps and synchronizing end devices' clocks at all times. However, we show that this approach is susceptible to a {\em frame delay attack} consisting of malicious frame collision and delayed replay. Real experiments show that the attack can affect the end devices in large areas up to about $50,000\,\text{m}^2$. In a broader sense, the attack threatens any system functions requiring timely deliveries of LoRaWAN frames. To address this threat, we propose a $\mathsf{LoRaTS}$ gateway design that integrates a commodity LoRaWAN gateway and a low-power software-defined radio receiver to track the inherent frequency biases of the end devices. Based on an analytic model of LoRa's chirp spread spectrum modulation, we develop signal processing algorithms to estimate the frequency biases with high accuracy beyond that achieved by LoRa's default demodulation. The accurate frequency bias tracking capability enables the detection of the attack that introduces additional frequency biases. Extensive experiments show the effectiveness of our approach.
\end{abstract}

\section{Introduction}
\label{sec:intro}
Low-power wide-area networks (LPWANs) enable direct wireless interconnections among end devices and gateways in geographic areas of square kilometers \cite{raza2017low}. It increases network connectivity as a defining characteristic of the Internet of Things (IoT). Among various LPWAN technologies (including NB-IoT and Sigfox), LoRaWAN \cite{LoRaWAN}, which is an open data link layer specification based on the LoRa modulation scheme \cite{lorawan-spec}, offers the advantages of using license-free ISM bands, low costs for end devices, and independence from managed cellular infrastructures.

LoRaWAN is promising for the applications of collecting low-rate monitoring data from geographically distributed sensors, such as utility meters, environment sensors, roadway detectors, industrial measurement devices, etc. 
All these applications require data timestamping as a basic system service, though they may require different timestamp accuracies. For instance, data center environment condition monitoring generally requires sub-second accuracy for sensor data timestamps to capture the thermodynamics \cite{chen2014sensor}.

Sub-second-accurate timestamps for the traffic data generated by roadway detectors can be used to reconstruct real-time traffic maps \cite{oh2002real}. In a range of industrial monitoring applications such as oil pipeline monitoring, milliseconds accuracy may be required \cite{pipeline}. In volcano monitoring, the onset times of seismic events detected by geographically distributed sensors require sub-10 milliseconds accuracy to be meaningful to volcanic earthquake hypocenter estimation \cite{liu2013volcanic}.

There are two basic approaches, namely, {\em sync-based} and {\em sync-free}, to data timestamping in wireless sensor networks (WSNs). In the sync-based approach, the sensor nodes keep their clocks synchronized and use the clock value to timestamp the data once generated. Differently, the sync-free approach uses the gateway with wall time to timestamp the data upon the arrival of the corresponding network packet. Based on various existing distributed clock synchronization protocols, multi-hop WSNs mostly adopt the sync-based approach. The sync-free approach is ill-suited for multi-hop WSNs, because the data delivery on each hop may have uncertain delays due to various factors such as channel contention among nodes.

In contrast, LoRaWANs prefer the sync-free approach for uplink data timestamping. Reasons are two-fold. First, different from multi-hop WSNs, LoRaWANs adopt a one-hop gateway-centered star topology that is free of the issue of hop-wise uncertain delays. Specifically, as the radio signal propagation time from an end device to the gateway is generally in microseconds, the LoRaWAN frame arrival time can well represent the time when the frame leaves the end device. As a result, timestamping the uplink data at the gateway can meet the milliseconds or sub-second timestamping accuracy requirements of many applications. Second, if the sync-based approach is adopted otherwise, the task of keeping the end devices' clocks synchronized at all times and the inclusion of timestamps in the LoRaWAN data frames will introduce communication overhead to the narrowband LoRaWANs. Therefore, performance-wise, the sync-free approach well matches LoRaWANs' star topology and addresses its bandwidth scarcity.

However, LoRaWAN's long-range communication capability also renders itself susceptible to wireless attacks that can be launched from remote and hidden sites. The attacks may affect many end devices in large geographic areas. In particular, the conventional security measures that have been included in the LoRaWAN specifications (e.g., frame confidentiality and integrity) may be inadequate to protect the network from wireless attacks on the physical layer. Therefore, it is of importance to study the potential wireless attacks against the sync-free data timestamping, since incorrect timestamps render sensor data useless and even harmful. For example, when applying LoRa for IoT object localization by triangulation, tiny timestamping error will lead to large localization errors. In this paper, we consider a basic threat of {\em frame delay attack} that directly invalidates the assumption of near-zero signal propagation time. Specifically, by setting up a {\em collider} device close to the LoRaWAN gateway and an {\em eavesdropper} device at a remote location, a combination of malicious frame collision and delayed replay may introduce arbitrary delays to the deliveries of uplink frames. Although wireless jamming and replay have been studied extensively, how easily they can be launched in a coordinated manner to introduce frame delay and how much impact (e.g., in terms of affected area) the attack can generate are still open questions in the context of LoRaWANs.

This paper answers these questions via real experiments. Our measurements show that LoRa demodulators have lengthy vulnerable time windows, in which the gateway cannot decode either the victim frame or the collision frame, and raises no alerts. Thus, it is easy to launch stealthy attacks by exploiting the vulnerable time windows. In particular, as the attack does not breach the integrity of the frame content and sequence, the attack cannot be solved by cryptographic protection and frame counting. Our experiments in a campus LoRaWAN show that, a fixed setup of a collider and an eavesdropper can subvert the sync-free data timestamping service for end devices in a large geographic area of about $50,000\,\text{m}^2$. In a broader sense, this attack threatens any system functions that require timely deliveries of uplink frames in LoRaWAN. Note that this attack is valid but marginally important in short-range wireless networks (e.g., Zigbee and Wi-Fi) because of the limited area affected by the attack and the difficulty in controlling the attack radios' timing. Differently, it is important to LoRaWANs because it can affect large geographic areas and the timing of the attack radios can be easily controlled due to LoRaWAN's long symbol times.

Therefore, an upgraded sync-free timestamping approach that integrates countermeasures against the attack and meanwhile preserves the bandwidth efficiency is desirable. Moreover, it should only require changes to the gateway. In this paper, we aim to develop awareness of the attack by monitoring the end devices' radio frequency biases (FBs), which are mainly caused by the manufacturing imperfections of the radio chips' internal oscillators. A deviation of FB detected by the gateway suggests the received frame may be a replayed one, since the adversary's replay device superimposes its own FB onto the replayed signal. To access the physical layer, we integrate a low-cost (US\$25 \cite{dongle-amazon}) software-defined radio (SDR) receiver \cite{rtl-sdr} with a commodity LoRaWAN gateway to form our LoRa TimeStamping ($\mathsf{LoRaTS}$) gateway.
We develop time-domain signal processing algorithms for $\mathsf{LoRaTS}$ to estimate the FB. Experiments show that (i) with a received signal-to-noise ratio (SNR) of down to $-18\,\text{dB}$, $\mathsf{LoRaTS}$ achieves an accuracy of $120\,\text{Hz}$ in estimating FB, which is just 0.14 parts-per-million (ppm) of the channel's central frequency of 869.75 MHz; (ii) the frame replay by an SDR transceiver introduces an additional FB of at least 0.24 ppm. Thus, $\mathsf{LoRaTS}$ can track FB to detect the replay step of the frame delay attack. In contrast, the LoRa's built-in FB estimation performed in the frequency domain \cite{eletreby2017empowering} does not achieve sufficient resolution to detect the attack. Note that the detection does not require uniqueness or distinctiveness of the FBs across different LoRa transceivers, because it is based on changes of FB.

The paper makes the following contributions:
\begin{itemize}
    \item We implement the frame delay attack against LoRaWAN. Simulations and experiments show the large sizes of the geographic areas vulnerable to the attack.
    \item Based on an analytic model of LoRa's chirp spread spectrum (CSS) modulation, we design a time-domain signal processing pipeline to accurately estimate end devices' FBs. The pipeline addresses challenges such as the need of microsecond-accurate arrival time estimation for the narrowband LoRa signal.
    \item Extensive experiments in both indoor and urban environments show that $\mathsf{LoRaTS}$ can detect the frame delay attacks that introduce additional FBs.
\end{itemize}

In summary, the feasibility of the attack and the large sizes of geographic areas vulnerable to the attack call for proper countermeasures. $\mathsf{LoRaTS}$, as a countermeasure, preserves the bandwidth efficiency of sync-free timestamping and requires no modifications on the LoRaWAN end devices. It is a low-cost countermeasure that increases the cost and technical barrier for launching effective frame delay attacks, since the attackers need to eliminate the tiny FBs of their radio apparatuses. Although completely solving the attack (including zero-FB attack and recovering from attack) still faces extra challenges, $\mathsf{LoRaTS}$ strikes a satisfactory trade-off between network efficiency and the security level required by typical LoRaWAN applications.

The rest of this paper is organized as follows. \sect\ref{sec:related} reviews related work; \sect\ref{sec:timestamping} describes sync-free data timestamping; \sect\ref{sec:security} studies the attack; \sect\ref{sec:lorasync} presents the $\mathsf{LoRaTS}$ design;
\sect\ref{sec:fingerprint} studies LoRa's FB and uses it to detect attack;
\sect\ref{sec:eval} presents experiment results;
\sect\ref{sec:discuss} discusses several issues;
\sect\ref{sec:conclude} concludes this paper.

\section{Related Work}
\label{sec:related}
Improving LoRaWAN's communication performance has received increasing research. Choir \cite{eletreby2017empowering} exploits the diverse FBs of the LoRaWAN end devices to disentangle colliding frames from different end devices. Choir uses the dechirping and Fourier transform processing pipeline to analyze FB, which does not provide sufficient resolution for detecting the tiny extra FB introduced by attack (see details in \sect\ref{subsubsec:extraction}). In this paper, based on an analytic model of LoRa's CSS modulation, we develop a new time-domain signal processing algorithm based on a least squares formulation to achieve the required resolution. Charm \cite{charm2018} exploits coherent combining to decode a frame from the weak signals received by multiple geographically distributed LoRaWAN gateways. It allows the LoRaWAN end device to use a lower transmitting power. Several recent studies \cite{peng2018plora,hessar19} have devised various backscatter designs for LoRa to reduce the power consumption of end devices. All the studies mentioned above focus on understanding and improving the data communication performance of LoRaWAN \cite{eletreby2017empowering,charm2018}, or reducing power consumption via backscattering \cite{peng2018plora,hessar19}. None of them specifically addresses efficient data timestamping, which is a basic system function of many LoRaWAN-based systems.

LongShoT \cite{ramirez2019longshot} is an approach to synchronize the LoRaWAN end devices with the gateway. Through low-level offline time profiling for a LoRaWAN radio chip (e.g., to measure the time delays between hardware interrupts and the chip's power consumption rise), LongShoT achieves sub-50 microseconds accuracy, which is echoed by our results on the accuracy of estimating signal arrival time using a different approach. LongShoT is designed for the LoRaWAN systems requiring tight clock synchronization. Differently, we address data timestamping and focus on the less stringent but more commonly seen milliseconds or sub-second accuracy requirements. Our sync-free approach releases the bandwidth from frequent clock synchronization operations.

Security of LoRaWAN is receiving research attention. In \cite{7985777}, Aras et al. discuss several possible attacks against LoRaWAN, including key compromise and jamming. The key compromise requires prior physical attack of memory extraction. In \cite{selective-jamming}, a selective jamming attack against certain receivers and/or certain application frames is studied. 
Different from the studies \cite{7985777, selective-jamming} that do not consider the stealthiness of jamming, we consider stealthy frame collision. From our results in \sect\ref{subsubsec:experiments}, the selective jamming in \cite{selective-jamming} cannot be stealthy because it cannot start jamming until the frame header is decoded and the corruption of payload must lead to integrity check failures. In \cite{robyns2017physical}, Robyns et al. apply supervised machine learning for end device classification based on the received LoRa signal. From our measurements, the dissimilarity between the original and the replayed signals is much lower than that among the original signals from different end devices. Thus, the supervised machine learning is not promising for attack detection. 

Device identification based on radiometric features has been studied for short-range wireless technologies. A radiometric feature is the difference between the nominal and the measured values of a certain modulation parameter. The work \cite{brik2008wireless} studied the radiometric features of IEEE 802.11 radios, including symbol-level features regarding signal magnitude and phase, as well as frame-level feature regarding carrier frequency. In LoRaWAN, the received signal strength is often rather low due to long-distance propagation or barrier penetration. As such, the signal magnitude radiometric feature cannot be used as a radiometric feature. As the phase of LoRa signal is arbitrary, it cannot be employed as a radiometric feature too. In this paper, we show that the bias of the LoRa signal's carrier frequency from the nominal value is an effective radiometric feature. This feature can be used to counteract the frame delay attack. Based on LoRa's CSS modulation, we develop a lightweight algorithm that can estimate this feature from the received LoRa signal. It requires a low-cost SDR receiver, unlike the expensive vector signal analyzer \cite{vectoranalyzer} used in \cite{brik2008wireless}.

\section{Data Timestamping in LoRaWAN}
\label{sec:timestamping}

\subsection{LoRaWAN Primer}
LoRa is a physical layer technique that adopts CSS modulation. LoRaWAN is an open data link specification based on LoRa. A LoRaWAN is a star network consisting of a number of {\em end devices} and a {\em gateway} that is often connected to the Internet. Gateways are often equipped with GPS receivers for time keeping. The transmission direction from the end device to the gateway is called {\em uplink} and the opposite is called {\em downlink}. LoRaWAN defines three classes for end devices, i.e., Class A, B and C. In Class A, each communication session must be initiated by an uplink transmission. There are two subsequent downlink windows. Class A end devices can sleep to save energy when there are no pending data to transmit. Class A adopts the ALOHA media access control protocol. Class B extends Class A with additional scheduled downlink windows. However, such scheduled downlink windows requires the end devices to have synchronized clocks, incurring considerable overhead as we will analyze shortly. Class C requires the end devices to listen to the channel all the time. Clearly, Class C is not for low-power end devices. In this paper, we focus on Class A, because it is supported by all commodity platforms and energy-efficient. To the best of our knowledge, no commodity platforms have out-of-the-box support for Class B that requires clock synchronization.

\subsection{Advantages of Sync-Free Timestamping}
\label{subsec:sync-vs-free}

Data timestamping, i.e., to record the {\em time of interest} in terms of the wall clock, is a basic system function required by the data collection applications for monitoring.
For a sensor measurement, the time of interest is the time instant when the measurement is taken by the end device. Multi-hop WSNs largely adopt the {\em sync-based} approach. Specifically, the clocks of the WSN nodes are synchronized to the global time using some clock synchronization protocol. Then, each WSN node can timestamp the data using its local clock. WSNs have to adopt this approach due primarily to that the multi-hop data deliveries from the WSN nodes to the gateway in general suffer uncertain delays. Thus, although the clock synchronization introduces additional complexity to the system implementation, it has become a standard component for systems requiring data timestamping. However, the clock synchronization introduces considerable communication overhead to the bandwidth-limited LoRaWANs.

We present an example to illustrate the overhead to maintain sub-10 milliseconds (ms) clock accuracy in LoRaWANs. Typical crystal oscillators in microcontrollers have drift rates of $30$ to $50\,\text{ppm}$ \cite{wizsync}. Without loss of generality, we adopt $40\,\text{ppm}$ for this example. With this drift rate, an end device needs 14 synchronization sessions per hour to maintain sub-$10\,\text{ms}$ clock accuracy. These 14 sessions represent a significant communication overhead for an end device. For instance, in Europe, a LoRaWAN end device adopting a spreading factor of 12 can only send 24 30-byte frames per hour to conform to the 1\% duty cycle requirement \cite{tech-LPWAN}. Although the synchronization information may be piggybacked to the data frames, a low-rate monitoring application may have to send the frames more frequently just to keep time. In addition, the data frames need to include data timestamps, each of which needs at least a few bytes. This is also an overhead given the bandwidth scarcity. 

To efficiently utilize LoRaWAN's scarce bandwidth and exploit its star topology, the sync-free timestamping approach can be adopted. In this approach, an end device transmits a sensor reading once generated. Upon receiving the frame, the gateway uses the frame arrival time as the data timestamp. The signal propagation time from the end device to the gateway, which is often microseconds, can be ignored for millisecond-accurate timestamping. Compared with the sync-based approach, this sync-free approach avoids the communication overhead caused by the frequent clock synchronization operations and the transmissions of timestamps. Thus, the sync-free approach is simple and provides bandwidth-saving benefit throughout the lifetime of the LoRaWANs.


\section{Security of Sync-Free Timestamping}
\label{sec:security}

The long-range communication capability of LoRaWAN enables the less complex and bandwidth-efficient sync-free timestamping. However, it may also be subjected to wireless attacks that can affect large geographic areas.
Having understood the benefit of sync-free timestamping, we also need to understand its security risk and the related countermeasure for achieving a more comprehensive assessment on the efficiency-security tradeoff. A major and direct threat against the sync-free approach is the {\em frame delay attack} that manipulates the frame delivery time to invalidate the assumption of near-zero signal propagation delay. We formally define the attack as follows.

\noindent {\bf Frame delay attack:} The end device and gateway are not corrupted by the adversary. However, the adversary may delay the deliveries of the uplink frames. The malicious delay for any uplink frame is finite. Moreover, the frame cannot be tampered with because of cryptographic protection.

The attack results in wrong timestamps under the sync-free approach. This section studies the attack implementation (\sect\ref{subsec:attack-implementation}), investigates the timing of malicious frame collision (\sect\ref{subsubsec:experiments}), and studies the size of the vulnerable area in which the end devices are affected by the attack (\sect\ref{subsec:attack-surface}).

\subsection{Attack Implementation}
\label{subsec:attack-implementation}
\subsubsection{Implementation steps}
\label{subsubsec:implementation-principle}

\begin{figure}
	\centering
	\includegraphics[width=.45\textwidth]{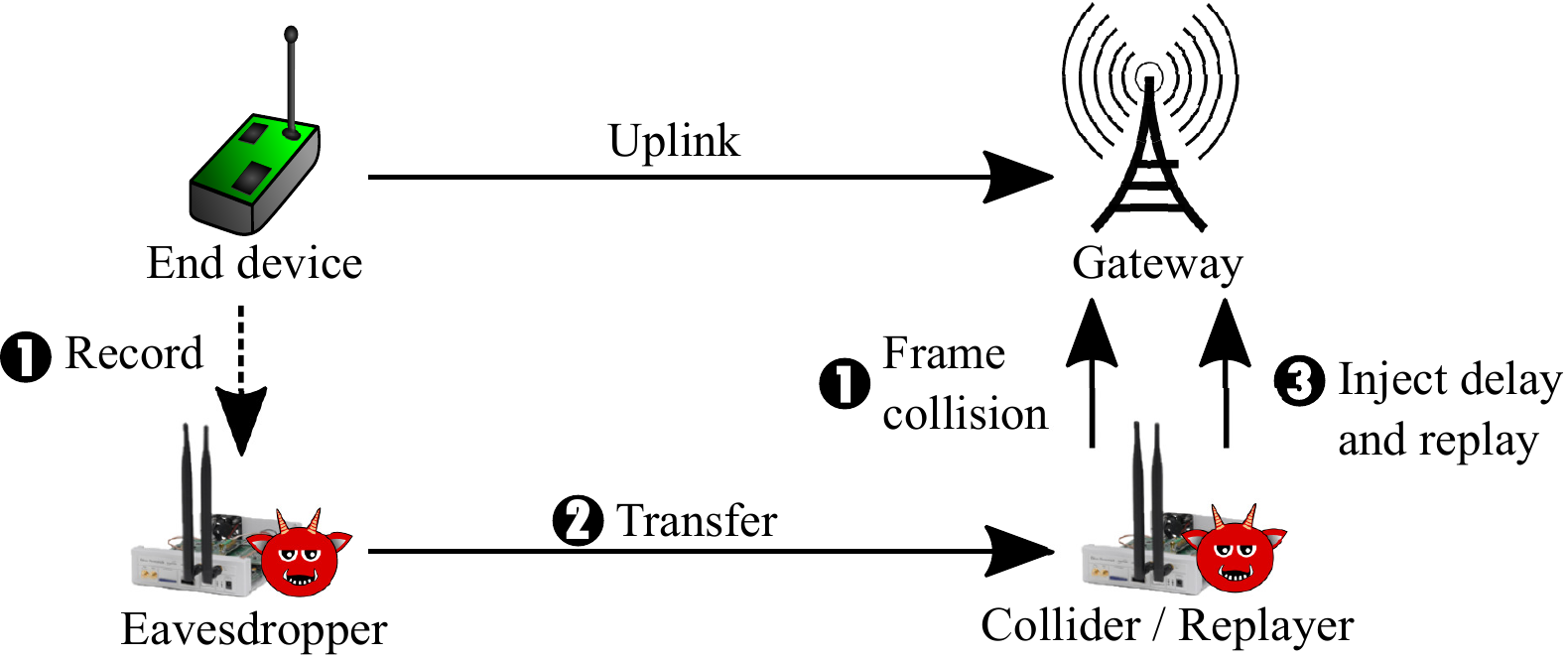}
	\caption{Steps for implementing frame delay attack.}
	\label{fig:lora_p2p_attack}
\end{figure}

Fig.~\ref{fig:lora_p2p_attack} illustrates the attack implementation. The adversary sets up two malicious devices called {\em eavesdropper} and {\em collider} that are close to the end device and the gateway, respectively. The attack consists of three steps. \ding{182} At the beginning, both the eavesdropper and the collider listen to the LoRa communication channel between the end device and the gateway. Once the collider detects an uplink frame transmission, it transmits a collision frame. In \sect\ref{subsubsec:experiments}, we will investigate experimentally a stealthy collision method such that the victim gateway does not raise any warning message to the application layer. Meanwhile, once the eavesdropper detects an uplink frame transmission, it records the radio waveform of the frame. Note that the collider may choose a proper transmitting power of the collision frame such that the collision can affect the victim gateway, while not corrupting the radio waveform recorded by the eavesdropper. \ding{183} The eavesdropper sends the recorded radio waveform data to the collider via a separate communication link that provides enough bandwidth (e.g., LTE). \ding{184} After a time duration of $\tau$ seconds from the onset time of the victim frame transmission, the collider replays the recorded radio waveform. Thus, in this paper, the {\em collider} and the {\em replayer} refer to the same attack device. The above collision-and-replay process does not need to decipher the payload of the recorded frame; it simply re-transmits the recorded radio waveform. As the gateway cannot receive the original frame and the integrity of the replayed frame is preserved, the gateway accepts the replayed frame even if it checks the cryptographically protected check sum and frame counter. The attack introduces a delay of $\tau$ seconds to the delivery of the frame.

We discuss several issues in the attack implementation. First, using a normal LoRaWAN frame to create malicious collision is more stealthy than brute-force jamming, since it may be difficult to differentiate malicious and normal collisions. Brute-force jamming can be easily detected and located. Second, as the adversary delays the uplink frame, how does the adversary know in time the direction of the current transmission? In LoRaWAN, the uplink preamble uses up chirps, whereas the downlink preamble uses down chirps. Thus, the adversary can quickly detect the direction of the current transmission within a chirp time. From our results in \sect\ref{subsubsec:experiments}, the collision should start after several chirps and before tens of chirps of the frame transmission. Thus, a time duration of one chirp for sensing the direction of the transmission does not impede the timeliness of the collision attack. Third, to increase the stealthiness of the replay attack, the replayer can well control the transmitting power of the replay such that only the victim gateway can receive the replayed frame. Fourth, the attack does not require clock synchronization between the eavesdropper and the collider.

\subsubsection{Discussion on a simple attack detector}
\label{subsubsec:discussion}

A simple attack detection approach is to perform round-trip timing and then compare the measured round-trip time with a threshold. However, this approach has the following three shortcomings. First, it needs a downlink transmission for each uplink transmission, which doubles the communication overhead. LoRaWAN is mainly designed and optimized for uplinks. For instance, a LoRaWAN gateway can receive frames from multiple end devices simultaneously using different spreading factors, whereas it can send a single downlink frame only at a time. This is because Class A specification requires that any downlink transmission must be unicast, in response to a precedent uplink transmission. Thus, the round-trip timing approach matches poorly with the uplink-downlink asymmetry characteristic of LoRaWAN. Second, with this simple attack detection approach, it is the end device detecting the attack after receiving the downlink acknowledgement. The end device needs to inform the gateway using another uplink frame that is also subject to malicious collision. Third, as the attacks are rare (but critical) events, continually using downlink acknowledgements to preclude the threat is a low cost-effective solution. In summary, this simple round-trip timing countermeasure is inefficient and error-prone.

\subsection{Timing of Malicious Frame Collision}
\label{subsubsec:experiments}
In this section, we study the timing of effective malicious frame collision. When investigating the geographic area affected by the attack, the ratio between the powers of the victim signal and the collision signal also needs to be considered. \sect\ref{subsec:attack-surface} will jointly consider the collision timing and the signal power ratio. We set up two SX1276-based LoRa nodes as the transmitter and the receiver, which are separated by about $5\,\text{m}$. We use a third LoRa node as the collider against the receiver. The distance between the collider and the receiver is about $1\,\text{m}$. Note that the Semtech SX1276 is the dominating 868MHz end device LoRa chip on the market. Although the quantified results obtained based on SX1276 may be chip specific, the qualitative results (i.e., the trend) presented below are consistent with our general understanding on wireless demodulation. Thus, the qualitative results provide general insights and implications. The gateway-class iC880A LoRaWAN concentrator and an open-source LoRa demodulator that we use in \sect\ref{subsec:attack-surface} also exhibit similar trend. In practice, the adversary may conduct experiments similar to those presented below to obtain the required attack timing once they know the model of the victim LoRa chip.

\begin{figure}
  \centering
  \includegraphics[width=.8\columnwidth]{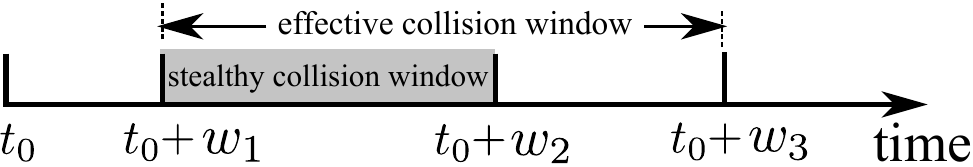}
  \caption{Collision attack time window.}
  \label{fig:time_window}
\end{figure}

From our experiments, there are three critical time windows (denoted by $w_1$, $w_2$, and $w_3$) after the onset time of the victim transmission (denoted by $t_0$). These time windows are illustrated in Fig.~\ref{fig:time_window}. If the onset time of the collision frame is in $[t_0, t_0 + w_1]$, the receiver most likely receives the collision frame only; if it is in $[t_0 + w_1, t_0 + w_2]$, the receiver receives neither frame and raises no alerts; if it is in $[t_0 + w_2, t_0 + w_3]$, the receiver reports ``bad frame'' and yields no frame content; if it is after $t_0 + w_3$, the receiver can receive both frames sequentially. Therefore, the time window $[t_0 + w_1, t_0 + w_2]$ is called {\em stealthy collision window} and the $[t_0 + w_1, t_0 + w_3]$ is called {\em effective collision window}. Note that we view the ``bad frame'' situation as effective attack, because the receiver cannot differentiate malicious and normal collisions based on the warning message.

We measure $w_1$, $w_2$, and $w_3$ under various settings for the spreading factor and the payload size of the victim frame. Table~\ref{table:attack-window} summarizes the results. From the results for $w_1$, the collision should start after the 5th chirp of the victim frame transmission. Explanation is as follows. (Note that as the demodulation mechanism of used SX1276 is proprietary and not publicly available, our explanations in this section are based on general understanding on wireless demodulation.) First, the receiver has not locked the victim frame's preamble until the 6th chirp. If the collision starts before the 5th chirp of the victim frame, the receiver will re-lock the collision frame's preamble with higher signal strength, resulting in reception of the collision frame. Second, the receiver locks the victim frame's preamble from the 6th chirp and simply drops any received radio data without reporting any error if any of the last three chirps (i.e., the 6th, 7th, and 8th chirps) of the preamble and/or the frame header are corrupted. For the latter case of frame header corruption, the radio chip cannot determine whether itself is the intended recipient and hence drops the received data. Thus, the collision should start after the 5th chirp of the victim frame.

We can also see that $w_2$ increases exponentially with the spreading factor. This is because: i) the total time for transmitting the preamble and frame header increases exponentially with the spreading factor; ii) corruption of the payload after the frame header leads to integrity check error and the ``bad frame'' message. The $w_3$ is roughly the time for transmitting the victim frame. Thus, if the collision onset time is after $t_0 + w_3$, both the victim and collision frames can be received.

\begin{table}
  \caption{Collision time windows for SX1276.}
  \label{table:attack-window}
  \centering
  \begin{tabular}{ccccccc}
    \toprule
    Spreading & Chirp & Preamble & Payload & $w_1$ & $w_2$ & $w_3$ \\
    factor $S$ & time & time & (byte) & & & \\
    \midrule
     & & & 10 & 5 & 28 & 141 \\
    7 & 1.024 & 8.2 & 20 & 5 & 38 & 156 \\
     & & & 30 & 6 & 41 & 165 \\
     & & & 40 & 6 & 54 & 178 \\
    \midrule
    7 & 1.024 & 8.2 & & 6 & 41 & 165 \\
    8 & 2.048 & 16.4 & 30 & 10 & 82 & 208 \\
    9 & 4.096 & 32.8 & & 22 & 156 & 274 \\
    \bottomrule
    \multicolumn{7}{l}{* Unit for chirp time, preamble time, $w_1$, $w_2$, $w_3$ is millisecond.} \\
  \end{tabular}
\end{table}

The above experiments show that, there is a time window of more than $20\,\text{ms}$ for the collision to corrupt the preamble partially and the frame header such that the victim simply drops the received data and raises no alerts. Collision starting in this window is stealthy. There is also an effective attack window of more than $100\,\text{ms}$. It is not difficult to satisfy such timing requirements using commodity radio devices.

\subsection{Size of Vulnerable Area}
\label{subsec:attack-surface}

In this section, through simulations and extensive experiments in a campus, we show that by setting up a collider and an eavesdropper at fixed locations, the frame delay attack can affect many end devices in a geographic area. The simulations based on realistic measurements with an open-source LoRa demodulator and a path loss model \cite{demetri2019automated} provide insights into understanding the vulnerable area. The experiments in the campus further capture other affecting factors such as terrain and signal blockage from buildings. In this section, the {\em core vulnerable area} refers to the geographic area in which the end devices are subject to stealthy collision and successful eavesdropping; the {\em vulnerable area} additionally includes the area in which the end devices are subject to the collision causing ``bad frame'' reports and successful eavesdropping.

\begin{figure}
  \begin{minipage}[t]{.24\textwidth}
    \includegraphics{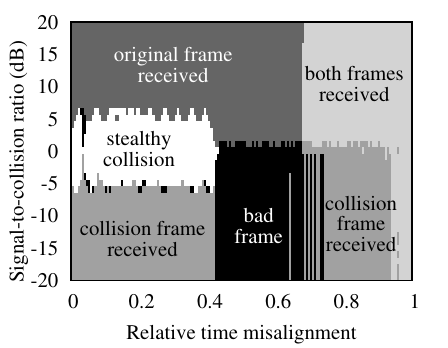}
    \caption{Result of \texttt{gr-lora}'s demodulation under collision with different signal-to-collision ratios and relative time misalignments.}
    \label{fig:diff_time_diff_power}
  \end{minipage}
  \hspace{0.5em}
  \begin{minipage}[t]{.2\textwidth}
    \includegraphics{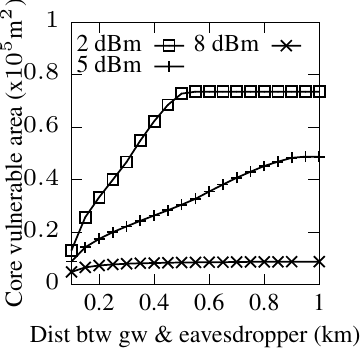}
    \caption{Core vulnerable area vs. distance between gateway and eavesdropper under various collision powers.}
    \label{fig:dist_area}
  \end{minipage}
\end{figure}

\begin{figure*}
  \centering
  \begin{minipage}[t]{.35\textwidth}
    \centering
    \includegraphics[height=2.4in, keepaspectratio]{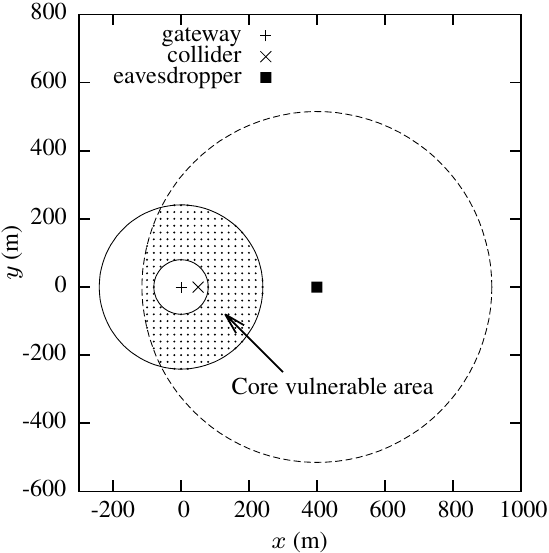}
    \vspace{-1em}
    \caption{The core vulnerable area (i.e., the shaded area) with gateway at $(0,0)$, collider at $(50,0)$, and eavesdropper at $(400,0)$. Collider's and victim end device's transmitting powers are $2\,\text{dBm}$ and $14\,\text{dBm}$, respectively. End devices in the ring centered at $(0,0)$ are subject to stealthy collision; end devices in the dashed circle are subject to successful eavesdropping.}
    \label{fig:ideal_attack_surface}
    \vspace{-1em}
  \end{minipage}
  \hspace{0.02\textwidth}
  \begin{minipage}[t]{0.60\textwidth}
    \vspace{-2.30in}
    \centering
    \includegraphics[width=\textwidth]{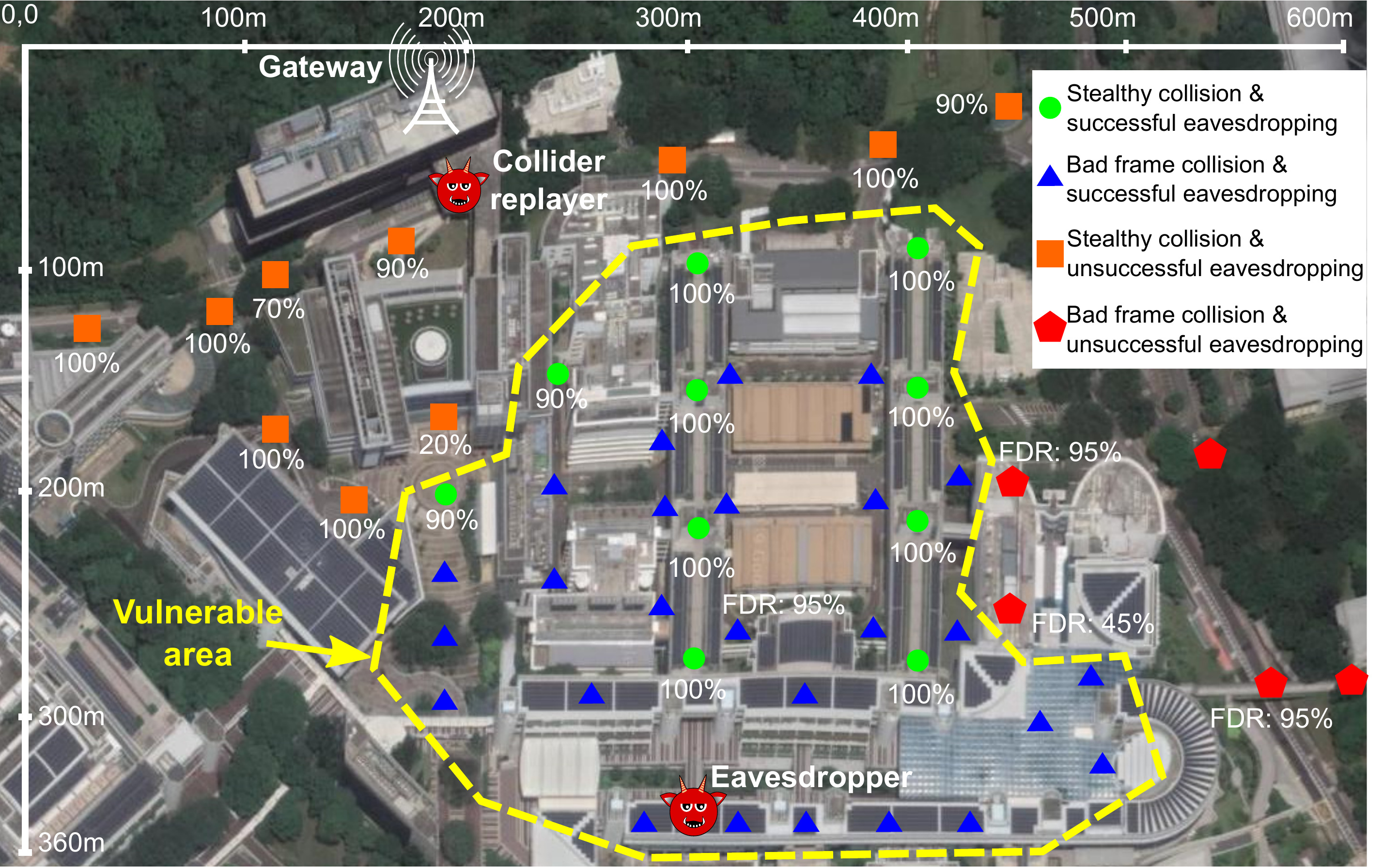}
    \vspace{-2em}
    \caption{Vulnerable area of a campus LoRaWAN. A gateway and a USRP-based eavesdropper are deployed on the rooftops of two buildings. A collider is deployed on an overhead bridge. We carry an end device to each of the marked locations and conduct an attack experiment. The four point shapes represent four types of attack outcomes. (Satellite image credit: Google Map)}
    \label{fig:real_experiment_detailed}
  \end{minipage}
\end{figure*}

\subsubsection{Simulations}
\label{subsubsec:simulation}

To study the vulnerable area, we need to consider the signal path loss and the ratio between the powers of the victim signal and the collision signal at the receiver. We call this ratio {\em signal-to-collision ratio} (SCR). To characterize attack timing, we define {\em relative time misalignment} (RTM) as $\frac{\text{collision time lag}}{\text{frame time}}$, where the collision time lag is the time lag of the collision onset from the victim signal onset.
In our simulation, the victim and collision frames have identical length but different payload contents. 
We generate the $I$ and $Q$ waveforms of these two frames using LoRa signal model.
We superimpose the two frames' signals to simulate collision. Moreover, we scale the amplitudes of the two signals and time-misalign them to create certain SCR and RTM. The sum signal is processed using an open-source LoRa demodulator \texttt{gr-lora} \cite{grlora}.
Fig.~\ref{fig:diff_time_diff_power} shows the demodulation results under various SCR and RTM settings.
We can see that if RTM is less than 0.4 and SCR at the gateway is within $[-6\,\text{dB}, 6\,\text{dB}]$, the collision is stealthy. The eavesdropped frame can be demodulated if SCR at the eavesdropper is greater than $6\,\text{dB}$.

We adopt a LoRa signal path loss model for urban areas proposed in \cite{demetri2019automated} based on real measurements. Specifically, the path loss $L$ in $\text{dB}$ is given by $L = 69.55 + 26.16\log f - 13.82\log h_b - (1.1\log f-0.7)h_m + (1.56\log f - 0.8) + (44.9 - 6.55\log h_b) \log d$, where the base of the logarithm is 10, $f$ is LoRa signal's central frequency in MHz, $h_m$ and $h_b$ are the heights of the transmitter and receiver in meters, and $d$ is the distance in kilometers between the transmitter and the receiver.
The frame delay attack is successful if the attacker can control RTM below 0.4 and satisfy the following two conditions:
\begin{align}
    -6\,\text{dB} &\leq P_{v} - L_{v,g} - (P_{c} - L_{c,g}) \leq 6\,\text{dB}, \label{eq:stealthy-condition} \\
    6\,\text{dB} &\leq P_{v} - L_{v,e} - (P_{c} - L_{c,e}), \label{eq:eavesdrop-condition}
    \vspace{-1em}
\end{align}
where the subscripts $v$, $g$, $c$, and $e$ respectively denote the victim end device, the gateway, the collider, and the eavesdropper; $P_x$ denotes the transmitting power of device $x$; $L_{x,y}$ denotes the path loss from device $x$ to $y$. Eq.~(\ref{eq:stealthy-condition}) is the condition for stealthy collision; Eq.~(\ref{eq:eavesdrop-condition}) is the condition for successful eavesdropping. The SCR thresholds of $6\,\text{dB}$ and $-6\,\text{dB}$ in Eqs.~(\ref{eq:stealthy-condition}) and (\ref{eq:eavesdrop-condition}) are from Fig.~\ref{fig:diff_time_diff_power}. Note that our modeling of successful eavesdropping in Eq.~(\ref{eq:eavesdrop-condition}) only considers the case that the signal from the collider at the eavesdropper has a power much higher than the noise floor, so that we can ignore the impact of noise on the eavesdropping.

Fig.~\ref{fig:ideal_attack_surface} shows an example of the areas defined by Eqs.~(\ref{eq:stealthy-condition}) and (\ref{eq:eavesdrop-condition}). The collider's and end device's transmitting powers are $2\,\text{dBm}$ and $14\,\text{dBm}$. The gateway's altitude is $25\,\text{m}$; the collider, eavesdropper, and end devices have an identical altitude of $0\,\text{m}$. As shown in Fig.~\ref{fig:ideal_attack_surface}, the ring centered at the gateway is defined by Eq.~(\ref{eq:stealthy-condition}); the disk area in the dashed circle is defined by Eq.~(\ref{eq:eavesdrop-condition}). Thus, the overlap between the ring and the disk is the core vulnerable area, which is $62,246\,\text{m}^2$.
Then, we vary the distance between the gateway and the eavesdropper (denoted by $d_{ge}$) and the $P_c$ setting. Fig.~\ref{fig:dist_area} shows the resulting core vulnerable area. We can see that the core vulnerable area in general increases with $d_{ge}$ and becomes flat after $d_{ge}$ exceeds a certain value. Moreover, among the three $P_c$ settings (i.e., 2, 5, and 8 dBm), $P_c = 2\,\text{dBm}$ gives larger core vulnerable areas. Reason of the above two observations is that the eavesdropper can achieve a larger eavesdropping area due to the weaker collision signal received by the eavesdropper. The core vulnerable area saturates because the eavesdropping area in the dashed circle illustrated in Fig.~\ref{fig:ideal_attack_surface} covers the entire ring area when $d_{ge}$ exceeds a certain value. Note that when $d_{ge}$ is very large, the noise power dominates and the core vulnerable area shrinks to zero.

The above simulation results suggest that the location of the gateway is the key information that the adversary needs to obtain. Based on that, the adversary can plan the placement of the collider and eavesdropper to affect a large geographic area. For the LoRaWANs adopting multiple gateways, the adversary can place a collider close to each of the gateways. In practice, the locations of the gateways can be obtained by the adversary in various ways (e.g., social engineering) and should not be relied on for the security of the system.

\subsubsection{Experiments in a campus LoRaWAN}
\label{subsubsec:exp_campus}
We conduct a set of experiments in an existing campus LoRaWAN to investigate the vulnerable area in real environments. Note that the LoRaWAN consists of three gateways that can cover the whole campus. Our experiments only involve one of the three gateways, which covers the area shown in Fig.~\ref{fig:real_experiment_detailed} that has a number of multistory buildings.
The gateway, which consists of an iC880a LoRaWAN concentrator board, a Raspberry Pi, and a high-gain antenna, is located on the rooftop of a building. Both the collider and the eavesdropper consist of a laptop computer and a USRP N210 each. The collider is placed on an overhead bridge attached to the gateway's building. The horizontal distance between the gateway and the collider is about $50\,\text{m}$. The eavesdropper is placed on the rooftop of another building that is about $320\,\text{m}$ from the gateway's building. We carry an SX1276-based LoRaWAN end device to each of the locations marked in Fig.~\ref{fig:real_experiment_detailed}, measure the frame delivery ratio (FDR), and perform an attack experiment. The measured FDRs at all the visited locations are 100\%, except the four locations labeled with non-100\% FDRs. Thus, the gateway can cover the accessible area shown in Fig.~\ref{fig:real_experiment_detailed}.

In each attack experiment, the end device's and the collider's transmitting powers are $14\,\text{dBm}$ and $8\,\text{dBm}$, respectively. All malicious collisions are effective. The outcomes can be classified into four categories, which are the combinations of the collision results (stealthy collision or ``bad frame'') and eavesdropping results (successful or unsuccessful). In Fig.~\ref{fig:real_experiment_detailed}, we use four point shapes to represent the four attack outcomes. The percentage below a location is the ratio of stealthy collisions. We can see that, at most locations close to the gateway and collider, the malicious collisions are stealthy. At the locations in the bottom most part of Fig.~\ref{fig:real_experiment_detailed}, the collisions cause gateway's bad frame reports. There is a transit region in the middle of Fig.~\ref{fig:real_experiment_detailed}, in which the collision outcomes are mixed. Note that the visited locations shown in Fig.~\ref{fig:real_experiment_detailed} are on the rooftops, in semi-outdoor corridors, or in indoor environments. The indoor/outdoor condition may affect the collision outcome type. At the locations in the area enclosed by the dashed polygon, the gateway can decode the frame that is recorded by the eavesdropper and then replayed by the collider, suggesting that the eavesdropping is successful. Thus, this area is the vulnerable area caused by the attack setup, which is about $50,000\,\text{m}^2$.

Note that the demodulation mechanism of the iC880a concentrator is proprietary and can be different from the open-source LoRa demodulator we used in \sect\ref{subsubsec:simulation}. The actual signal propagation behaviors in the campus LoRaWAN can be much more complex than the model used in \sect\ref{subsec:attack-surface}. However, the simulation result (Fig.~\ref{fig:ideal_attack_surface}) and real experiment result (Fig.~\ref{fig:real_experiment_detailed}) show similar patterns, i.e., the eavesdropping area is around the eavesdropper and the core vulnerable area is a belt region between the gateway and the eavesdropper. Thus, our modeling and simulations in \sect\ref{subsec:attack-surface} provide useful understanding on the LoRaWAN vulnerability.


\subsection{Attack-Aware Sync-free Timestamping}
\label{subsec:based-vs-free}

The results in \sect\ref{subsubsec:experiments} and \sect\ref{subsec:attack-surface} have shown that the frame delay attack is a real and immediate concern for LoRaWAN. Moreover, a fixed setup of a collider and an eavesdropper can subvert the sync-free timestamping and in a broader sense, any system functions requiring timely frame deliveries, for many end devices in a large geographic area. From \sect\ref{subsec:sync-vs-free}, sync-free timestamping has a main advantage of lower bandwidth usage. This advantage takes effect throughout the lifetime of the network. Therefore, if we can devise a low-overhead countermeasure for sync-free timestamping against the attack that may rarely occur (but can be devastating once occurred), we can enjoy the continuing and important advantage of lower bandwidth usage and strike a good trade-off between performance and security. Ideally, this countermeasure runs at the gateway only and does not require any modifications to the hardware and software of the end devices. In this paper, we aim to develop awareness of the frame delay attack by detecting the replay step of the attack. With this awareness, the sync-free timestamping will not be misled unknowingly. To this end, we present $\mathsf{LoRaTS}$ and its attack detection approach in the next two sections.


\section{$\mathsf{LoRaTS}$ Gateway}
\label{sec:lorasync}
\subsection{$\mathsf{LoRaTS}$ Gateway Hardware}
\label{subsec:hardware}

\begin{figure}
  \begin{minipage}[t]{.22\textwidth}
    \includegraphics[width=\textwidth]{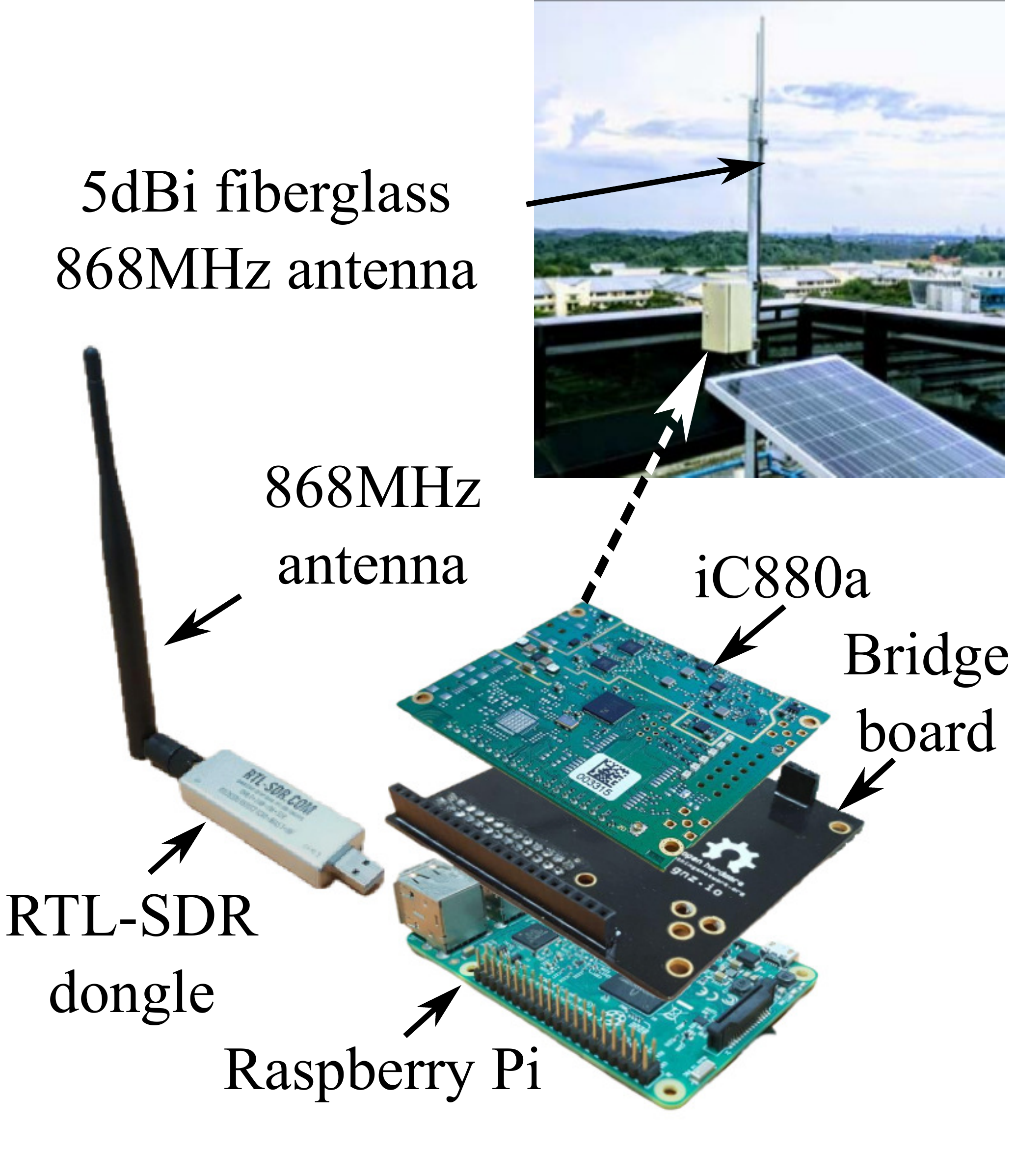}
    \caption{$\mathsf{LoRaTS}$ hardware prototype consisting of Raspberry Pi, iC880a concentrator, bridge board, RTL-SDR USB dongle.}
    \label{fig:softlora_gateway}
  \end{minipage}
  \hfill
  \begin{minipage}[t]{.24\textwidth}
    \includegraphics{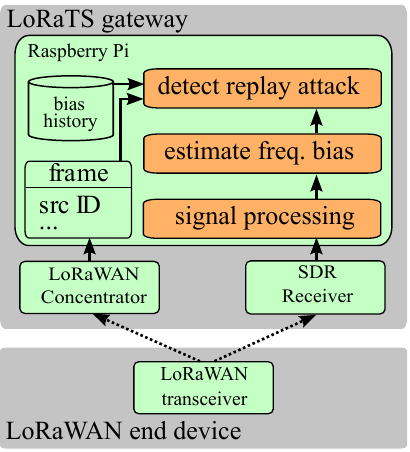}
    \caption{$\mathsf{LoRaTS}$ software. Bottom part is end device; upper part is gateway; solid arrows are local data flows; dashed arrows are transmissions.}
    \label{fig:lorawan_arch}
  \end{minipage}
\end{figure}

To detect the attack, we integrate an SDR receiver with a LoRaWAN gateway to monitor the physical layer. In this paper, we use RTL-SDR USB dongles based on the RTL2832U chipset \cite{rtl-sdr}, which were originally designed to be DVB-T TV tuners. It is cheap (US\$25 only) and covers the LoRaWAN bands. It can operate at $2.4\,\text{Msps}$ reliably for extended time periods. Thus, the sampling resolution is $1/2.4\,\text{Msps} = 0.42\,\mu\text{s}$. Our research is conducted based on a $\mathsf{LoRaTS}$ hardware prototype that integrates a Raspberry Pi, an iC880a LoRaWAN concentrator, and an RTL-SDR USB dongle. Fig.~\ref{fig:softlora_gateway} shows the prototype. An $868\,\text{MHz}$ antenna is used with the RTL-SDR to improve signal reception.

The SDR receiver is used to capture the radio signal over a time duration of the first two preamble chirps of an uplink frame. The first sampled chirp is used to determine the signal's arrival timestamp, whereas the second sampled chirp is used to estimate the FB of the transmitter. The accurate timestamp is a prerequisite of the FB estimation. As only two chirps' radio waveform is analyzed, the Raspberry Pi suffices for performing the computation. Instead of using RTL-SDR, a full-fledged SDR transceiver (e.g., USRP) can be used to design a customized gateway with physical layer access. However, this design loses the factory-optimized hardware-speed LoRa demodulation built in the iC880a concentrator. Moreover, full-fledged SDR transceivers are often 10x more expensive than $\mathsf{LoRaTS}$. The low-cost, low-power, listen-only RTL-SDR suffices for developing the attack detector.

\subsection{$\mathsf{LoRaTS}$ Gateway Software}
\label{subsec:application}

The upper part of Fig.~\ref{fig:lorawan_arch} illustrates the software architecture of $\mathsf{LoRaTS}$ to detect the attack. It is based on the results in the subsequent sections of this paper. The uplink transmission from the end device is captured by both the gateway's LoRaWAN concentrator and the SDR receiver. The LoRaWAN concentrator demodulates the received radio signal and passes the frame content to the Raspberry Pi. Signal processing algorithms are applied on the LoRa signal after down-conversion by the SDR receiver to determine precisely the arrival time of the uplink frame, estimate the transmitter's FB, and detect whether the current frame is a replayed one. The replay detection is by checking whether the estimated FB is consistent with the historical FBs associated with the transmitter ID contained in the current frame. Thus, the gateway is aware of the attack and can take necessary actions. Note that $\mathsf{LoRaTS}$ uses the SDR receiver to obtain FBs, rather than to decode the frame.

We use an Akaike Information Criterion (AIC) \cite{sleeman1999robust} based algorithm to accurately detect the onset time of the received LoRa frame and locate chirps. The root-mean-square deviation (RMSD) of AIC's onset time detection error is less than 5 $\mu s$ when the SNR is down to -20 dB \cite{lora-sync-report2}. Thus, AIC achieves robust onset time detection in the presence of strong noises.


\section{Frame Delay Attack Detection}
\label{sec:fingerprint}

Internal oscillators for generating carriers generally have FBs due to manufacturing imperfection. This section develops algorithms for estimating LoRa transmitters' FBs based on LoRa's CSS modulation and use them to detect the frame delay attack. Note that the existing FB estimation algorithms developed for other radios cannot be ported to LoRa due to different modulation schemes. For instance, the FB estimation for OFDM \cite{yao2005blind} is apparently not applicable for LoRa CSS. As discussed later, LoRa demodulation's built-in FB estimation technique does not provide sufficient resolution. Thus,  highly accurate FB estimation for LoRa CSS is a non-trivial problem.

\subsection{FB Estimation}
\label{subsubsec:extraction}

\begin{figure}
  \begin{minipage}[t]{.24\textwidth}
    \centering 
    \includegraphics[width=\textwidth, height=2.2cm, keepaspectratio]{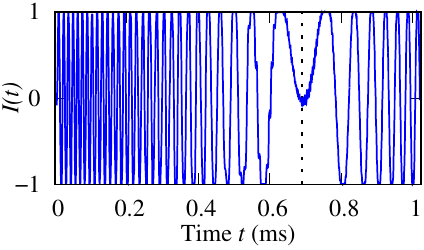}
    \caption{Actual $I$ data of a preamble chirp. The time shift of the axis of symmetry represented by the dashed line is caused by FB.}
    \label{fig:I_signal}
  \end{minipage}
  \begin{minipage}[t]{.20\textwidth}
    \includegraphics[width=\textwidth, height=2.2cm]{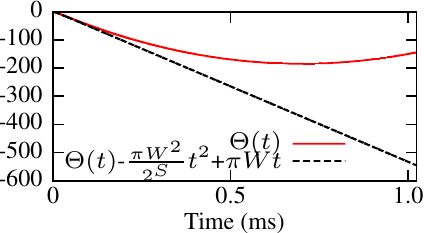}
    \caption{FB estimation using linear regression. The two curves are generated from actual $I$ and $Q$ data.}
    \label{fig:extraction2}
  \end{minipage}
\end{figure}

This section describes algorithms for estimating the transmitter's FB based on an up chirp in the preamble.
First, we analyze the impact of the transmitter's and SDR receiver's FBs (denoted by $\delta_{\mathrm{Tx}}$ and $\delta_{\mathrm{Rx}}$) on the $I$ and $Q$ traces.
The up chirp's instantaneous frequency accounting for $\delta_{\mathrm{Tx}}$ is $f(t) = \frac{W^{2}}{2^{S}} \cdot t - \frac{W}{2} + f_c + \delta_{\mathrm{Tx}}$, $t \in \left[ 0, \frac{2^S}{W} \right]$. The two local unit-amplitude orthogonal carriers generated by the SDR receiver are $\sin(2 \pi (f_c + \delta_{\mathrm{Rx}})t + \theta_{\mathrm{Rx}})$ and $\cos(2 \pi (f_c + \delta_{\mathrm{Rx}})t + \theta_{\mathrm{Rx}})$. After mixing and low-pass filtering, the $I$ and $Q$ components of the received up chirp can be derived as $I(t) = \frac{A(t)}{2} \cos \Theta(t)$ and $Q(t) = \frac{A(t)}{2} \sin \Theta(t)$, where the angle $\Theta(t)$ is given by
\begin{equation*}
  \Theta(t) = \frac{\pi W^2}{2^S} t^2 - \pi W t + 2\pi \delta t + \theta_{\mathrm{Tx}} - \theta_{\mathrm{Rx}}, \quad \delta = \delta_{\mathrm{Tx}} - \delta_{\mathrm{Rx}}. \inlineeqnum\label{eq:Theta-with-bias}
\end{equation*}
When $\delta=0$, the axis of symmetry of $I(t)$ is located at the midpoint of the preamble chirp time. As shown in Fig.~\ref{fig:I_signal}, a negative $\delta$ causes a right shift of the axis of the symmetry in the time domain, whereas a positive $\delta$ causes a left shift.

For a certain SDR receiver, the FB estimation problem is to estimate $\delta$ from the captured $I$ and $Q$ traces. We do not need to estimate $\delta_{\mathrm{Tx}}$, because for a certain SDR receiver with a nearly fixed $\delta_{\mathrm{Rx}}$, a change in $\delta$ indicates a change in $\delta_{Tx}$ and a replay attack. In fact, FB estimation is a prerequisite of LoRa demodulation. Now, we discuss the incompetence of the LoRa demodulators' built-in FB estimation technique for attack detection. LoRa's CSS scheme evenly divides the whole channel bandwidth of $W\,\text{Hz}$ into $2^{S}$ bins, where $S$ is the spreading factor. The starting frequency of a bin corresponds to a symbol state. Since the preamble chirp linearly swaps the channel bandwidth, its starting frequency can be viewed as the FB. LoRa demodulation firstly applies {\em dechirping} and then FFT to identify the preamble's and any data chirp's starting frequency bin indexes. The difference between the two indexes is the symbol state. As FFT achieves a resolution of $\frac{1}{x}\,\text{Hz}$ using $x$ seconds of data, the Fourier transform of a chirp with length of $\frac{2^S}{W}$ seconds has a frequency resolution of $\frac{W}{2^S}\,\text{Hz}$. This is also the resolution of the built-in FB estimation. Thus, for low spreading factor settings, the resolution may be poor. For instance, when $S=7$ and $W=125\,\text{kHz}$, the resolution is $976.56\,\text{Hz}$. However, as we will show in \sect\ref{subsec:countermeasures}, this near-$1\,\text{kHz}$ resolution is insufficient to detect attacks that introduce sub-$1\,\text{kHz}$ FBs. The colliding frame disentanglement approach Choir \cite{eletreby2017empowering} also uses the dechirping-FFT pipeline to analyze FB. Thus, it is subject to the insufficient resolution. To achieve higher resolutions, this section presents two time-domain approaches designed based on Eq.~(\ref{eq:Theta-with-bias}).

\subsubsection{Linear regression approach}
\label{subsubsec:linear-regression}
Eq.~(\ref{eq:Theta-with-bias}) can be rewritten as $\Theta(t) - \frac{\pi W^2}{2^S} t^2 + \pi W t = 2\pi \delta t + \theta$, which is a linear function of $t$ with $2\pi\delta$ as the slope. Thus, the slope can be estimated by linear regression based on the data pairs $(t, \Theta(t) - \frac{\pi W^2}{2^S} t^2 + \pi W t)$, where $t \in \left[ 0, \frac{2^S}{W} \right]$, $\Theta(t) = \mathrm{atan2}(Q(t), I(t)) + 2k\pi$, and $k \in \mathbb{Z}$ rectifies the multi-valued inverse tangent function $\mathrm{atan2}(\cdot, \cdot) \in (-\pi, \pi)$ to an unlimited value domain. The details of the rectification are omitted here due to space limitation and can be found in \cite{lora-sync-report2}.
Note that the $I(t)$ and $Q(t)$ are the $I$ and $Q$ data traces captured by the SDR receiver for a complete preamble chirp. The preamble onset time detected by AIC is used to segment the $I$ and $Q$ traces to chirps. 
Fig.~\ref{fig:extraction2} shows the $\Theta(t)$ computed from real $I$ and $Q$ traces of the second chirp of a preamble emitted by an SX1276-based end device and captured by $\mathsf{LoRaTS}$'s SDR receiver. It also shows $\Theta(t) - \frac{\pi W^2}{2^S} t^2 + \pi W t$, which is indeed a linear function of time.
As the linear regression approach has a closed-form formula to compute $\delta$, it has a complexity of $\mathcal{O}(1)$.

\subsubsection{Least squares approach}
\label{subsec:low-snr}
The LoRa signals can be very weak after long-distance propagation or barrier penetration. The LoRa's demodulation is designed to address low SNRs.
For SX1276, the minimum SNRs required for reliable demodulation with spreading factors of 7 to 12 are $-7.5\,\text{dB}$ to $-20\,\text{dB}$ \cite{sx1276-data}. We aim at extracting FB at such low SNRs. We solve a least squares problem: $\argmin_{\theta_{\mathrm{Tx}} - \theta_{\mathrm{Rx}} \in [0, 2\pi), \delta} \sum_{t \in \left[ 0, 2^S/W \right]} \left( Q(t) - A \sin \Theta(t) \right)^2 + \left( I(t) - A \cos \Theta(t) \right)^2$,
where $Q(t)$ and $I(t)$ are the received $Q$ and $I$ traces; $\Theta(t)$ is given by Eq.~(\ref{eq:Theta-with-bias}); $A \sin \Theta(t)$ and $A \cos \Theta(t)$ are the noiseless $Q$ and $I$ templates. The above formulation requires that the $Q$ and $I$ templates have an identical and constant amplitude $A$. As the second preamble chirp can meet this requirement, we use it for FB estimation. The $A$ can be estimated as the square root of the difference between the average powers of the LoRa signal and the pure noise. We use a \texttt{scipy} implementation of the differential evolution algorithm to solve the least squares problem. Raspberry Pi uses 0.69 seconds to solve it.

\subsubsection{Performance comparison}

\begin{figure}
  \subfigure[Linear regression]
  {
    \includegraphics{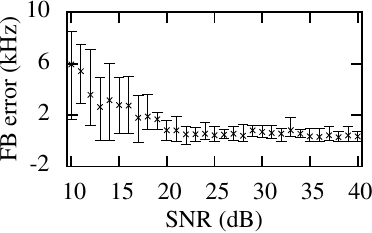}
    \label{fig:lr_estimation_error}
  }
  \subfigure[Least squares]
  {
    \includegraphics{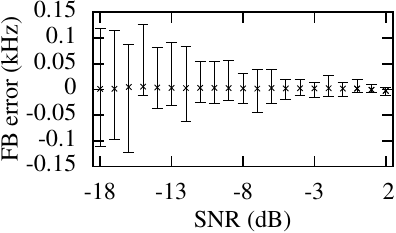}
    \label{fig:ls_estimation_error}
  }
  \caption{FB estimation errors vs. SNR.}
  \label{fig:fb-errors}
\end{figure}

We compare the FB estimation accuracy of the linear regression and the least squares approaches.
Fig.~\ref{fig:fb-errors} shows the results. For each SNR setting, 20 LoRa $I$ and $Q$ traces with random FBs are generated using the signal model in Eq.~(\ref{eq:Theta-with-bias}). We also generate 20 noises traces; the magnitude of the noise is controlled to achieve the specified SNR. In Fig.~\ref{fig:fb-errors}, each error bar showing the 20\%- and 80\%-percentiles is from the 20 FB estimation results performed on the sum signals of the generated ideal LoRa signals and noise. From Fig.~\ref{fig:lr_estimation_error}, the linear regression approach can achieve low FB estimation errors when the SNR is very high (e.g., $40\,\text{dB}$). However, it performs poorly for low SNRs. This is caused by the susceptibility of the inverse tangent rectification to noises. Specifically, as the inverse tangent rectification is based on a heuristic to detect $\mathrm{atan2}$'s sudden transitions between $-\pi$ and $\pi$, large noises leads to false positive detection of the transitions.
Differently, the least squares approach maintains the FB estimation error within $120\,\text{Hz}$ (i.e., $0.14\,\text{ppm}$), when the SNR is down to $-18\,\text{dB}$. Thus, the rest of this paper adopts the noise-resilient least squares approach, though it is more compute-intensive.

\subsubsection{FB measurements for 16 end devices}
\label{subsubsec:fb_indoor}
\begin{figure}
  \centering
  \includegraphics{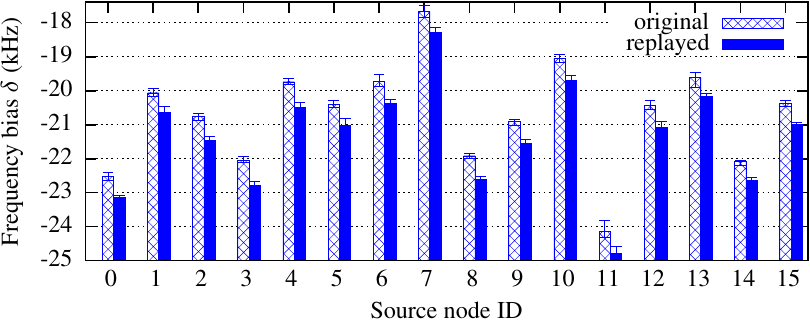}
  \caption{FBs estimated from the original LoRa signals from 16 end nodes and those replayed by a USRP-based replayer. The error bar shows mean, minimum, and maximum of FBs in 20 frame transmissions.}
  \label{fig:frequency_bias}
\end{figure}

We use an RTL-SDR to estimate the FBs of 16 SX1276-based end devices. In each test for an end device, the distance between the end device and the RTL-SDR is about $5\,\text{m}$. The error bars labeled ``original'' in Fig.~\ref{fig:frequency_bias} show the results. We can see that the FBs for a certain node are stable and the nodes generally have different FBs. The absolute FBs are from $17\,\text{kHz}$ to $25\,\text{kHz}$, which are about $20\,\text{ppm}$ to $29\,\text{ppm}$ of the nominal central frequency of $869.75\,\text{MHz}$. Some nodes have similar FBs, e.g., Node 3, 8, and 14. Note that the detection of the replay attack is based on the fact that the replayed transmission has a different FB. In other words, the attack detection does not require distinct FBs among different end devices. From Fig.~\ref{fig:frequency_bias}, we also observe that all nodes have negative FB measurements, which means that $\delta_{\mathrm{Tx}} < \delta_{\mathrm{Rx}}$, where $\delta_{\mathrm{Tx}}$ and $\delta_{\mathrm{Rx}}$ are the unknown FBs of the end device and the RTL-SDR. Note that as the RTL-SDR is a low-cost device, it may have a large FB causing the negative relative FB measurements.

\subsection{Replay Attack Detection}
\label{subsec:countermeasures}

The replayer also has an FB. The error bars labeled ``replayed'' in Fig.~\ref{fig:frequency_bias} show the FBs estimated from the LoRa signals received by the $\mathsf{LoRaTS}$'s SDR receiver when a USRP replays the radio waveform captured by itself in the experiments presented in \sect\ref{subsubsec:extraction}.
Compared with the results labeled ``original'', the FBs of the replayed transmissions are consistently lower. This is because the USRP has a negative FB.
The average additional FBs introduced by the replayer range from $-543$ to $-743\,\text{Hz}$, i.e., $0.62$ to $0.85\,\text{ppm}$  of the channel's central frequency. Thus, with the FB estimation accuracy of $0.14\,\text{ppm}$ achieved under low SNRs in \sect\ref{subsec:low-snr}, the additional FBs caused by the replay attack can be detected.

Based on the above observation, we describe an approach to detect the delayed replay. $\mathsf{LoRaTS}$ maintains a database of the FBs of the nodes with which it communicates. This database can be built offline or at run time using its SDR receiver in the absence of attacks. To address the end devices' time-varying radio frequency skews due to run-time conditions like temperature, $\mathsf{LoRaTS}$ can continuously update the database entries based on the FBs estimated from recent frames. To decide whether the current received frame is a replayed frame, the $\mathsf{LoRaTS}$ gateway checks whether the FB of the current received frame is within the acceptable FB range of the end device based on the database. This detection approach is applied after the $\mathsf{LoRaTS}$ gateway decodes the frame to obtain the end device ID. The FB estimated from a frame detected as a replayed one should not be used to update the database.

This detection mechanism forms a first line of defense against the frame delay attacks that introduce extra FBs. It gives awareness of the attack that is based on the logistics of collision and record-and-replay. With knowledge of our detector, the attackers may invest more resources and efforts to hide their radiometrics. \sect\ref{sec:discuss} will discuss potential approaches to eliminate the extra FBs. While this attack-defense chase is interesting, in this paper, we focus on showing the vulnerability of sync-free timestamping and propose the FB-based attack detector that forces the attackers to hide their radiometrics with increased cost and technical barriers.


\section{Experiments}
\label{sec:eval}
\subsection{Experiments in a Multistory Building}
\label{sec:indoor-exp}

\begin{figure}[t!]
  \centering
  \includegraphics{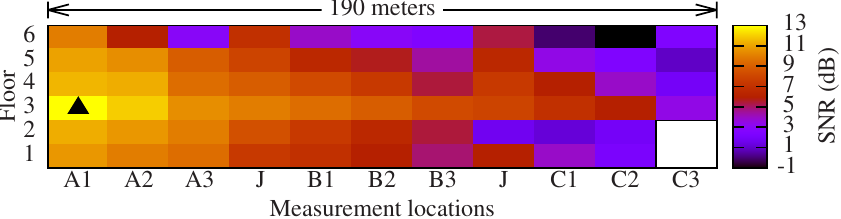}
  \caption{SNR survey in a building (lateral view) with 3 sections (A, B, C) and 2 junctions (J). The triangle represents the fixed node.}
  \label{fig:snr-survey}
\end{figure}

LoRaWAN can be used for indoor applications, such as utility metering. We conduct a set of experiments to investigate the feasibility of attack and effectiveness of our attack detector in a concrete building with six floors. The building has three sections and two section junctions along its long dimension of 190 meters. Fig.~\ref{fig:snr-survey} illustrates a lateral view of the building. First, we survey the SNR inside the building to understand the signal attenuation. We deploy a fixed LoRaWAN transmitter in Section~A on the 3rd floor. Then, we carry an SDR receiver to different positions inside the building to measure the SNR. 
In each section, we measure three positions. The heat map in Fig.~\ref{fig:snr-survey} shows the SNR measurements. We can see that the SNR decays with the distance between the two nodes. The SNRs are from $-1\,\text{dB}$ to $13\,\text{dB}$.
Then, we conduct the following experiments. By default, we set $S = 12$.

\noindent {\bf Attack experiments:}
We deploy an iC880a-based gateway and an SX1276-based end device in Section A1 of the 3rd floor and Section C3 of the 6th floor, respectively.
The LoRa signals are significantly attenuated after passing through multiple building floors. If the end device adopts a spreading factor of 7, it cannot communicate with the gateway. A minimum spreading factor of 8 is needed for communications.
We deploy two USRP N210 stations as the eavesdropper and the collider, next to the end device and the gateway, respectively.
We set the transmitting power of the end device and the collider to be $14\,\text{dBm}$.
The malicious collision is stealthy to the gateway; the eavesdropping is successful.
Thus, the frame delay attack can be launched in this building.

\begin{figure}[t]
  \centering
  \includegraphics{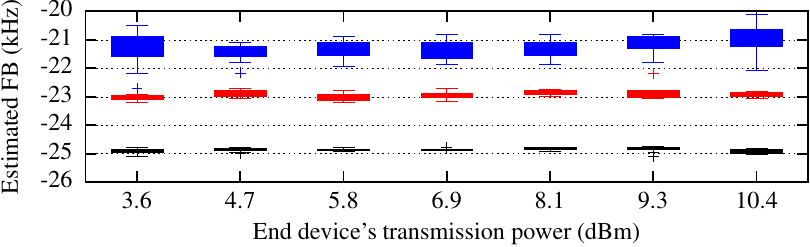}
  \caption{Estimated FB vs. transmitting power of the end device. Each box plot shows min, max, 25\% and 75\% percentiles. (1) Bottom row: end device to eavesdropper; (2) Mid row: end device to $\mathsf{LoRaTS}$ gateway; (3) Top row: replayer to $\mathsf{LoRaTS}$.}
  \label{fig:delta-power}
\end{figure}

\noindent {\bf Impact of transmitting power on FB estimation:} Fig.~\ref{fig:delta-power} shows the estimated FBs versus the end device's transmitting power under different settings. The bottom row of black box plots are the FBs estimated by the eavesdropper when the end device transmits the uplink frame with different transmitting powers. The middle row of red box plots are the FBs estimated by the $\mathsf{LoRaTS}$ gateway in the absence of the frame collision and replay attacks. Thus, the FBs estimated by the eavesdropper and the $\mathsf{LoRaTS}$ gateway are different. This is because that as analyzed in \sect\ref{subsubsec:extraction}, the estimated FB $\delta$ contains the transmitter's and receiver's FBs $\delta_{\mathrm{Tx}}$ and $\delta_{\mathrm{Rx}}$. Note that the eavesdropper and the $\mathsf{LoRaTS}$ gateway in general have different FBs. From Fig.~\ref{fig:delta-power}, the end device's transmitting power has little impact on the FB estimation.

\noindent {\bf Additional FB introduced by replayer:}
In Fig.~\ref{fig:delta-power}, the top row of blue box plots are the FBs estimated by the $\mathsf{LoRaTS}$ when the replayer replays the radio waveform recorded by the eavesdropper.
When the end device adopts a higher transmitting power, the replayed signal also has higher power. By comparing the middle and the top rows, we can see that the replay attack introduces an additional FB of about $2\,\text{kHz}$, which is $2.3\,\text{ppm}$ of the LoRa channel's central frequency. Therefore, the FB monitoring can easily detect the replay attack. Compared with the results in Fig.~\ref{fig:frequency_bias} showing additional FBs of $0.62$ to $0.85\,\text{ppm}$, the FBs in this set of experiments are higher. This is because that here we use two different USRPs as the eavesdropper and replayer; their FBs are superimposed.

\subsection{Outdoor Experiments with Longer Distance}
\begin{figure}[t]
  \centering
  \includegraphics{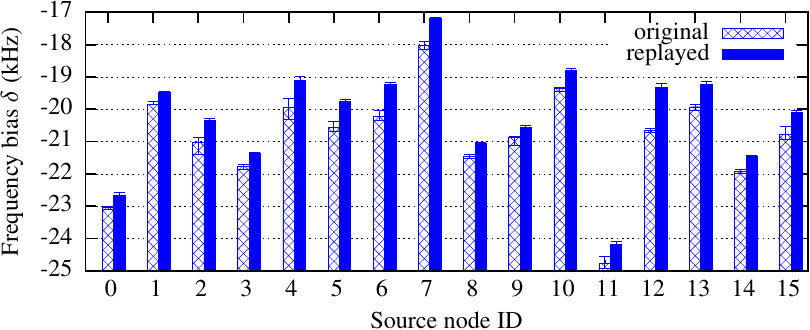}
  \caption{$\mathsf{LoRaTS}$'s FB estimates when the distances between the gateway and the end devices are about $1.07\,\text{km}$.}
  \label{fig:long_dist_fb}
\end{figure}

We deploy SX1276-based end devices in an outdoor parking lot. We replace the iC880a-based gateway shown in Fig.~\ref{fig:real_experiment_detailed} with a $\mathsf{LoRaTS}$ gateway. The distance between the end device and the $\mathsf{LoRaTS}$ gateway is about $1.07\,\text{km}$.
The collider shown in Fig.~\ref{fig:real_experiment_detailed} is also used in this set of experiments. The eavesdropper is deployed at a location about $200\,\text{m}$ from the end device. When the transmitting powers of the end device and the collider are $14\,\text{dBm}$ and $8\,\text{dBm}$, respectively, we can successfully launch the frame delay attack.
Then, we investigate the additional FBs introduced by the replay attack. Fig.~\ref{fig:long_dist_fb} shows $\mathsf{LoRaTS}$'s FB estimates for the frames transmitted by 16 end devices and the corresponding replays.
The extra FBs introduced by the attack is up to 1.76 ppm. Thus, the attack can be detected.

\subsection{Temporal Stability of FB}

\begin{figure}
  \subfigure[FB of an device \& temperature.]
  {
    \includegraphics{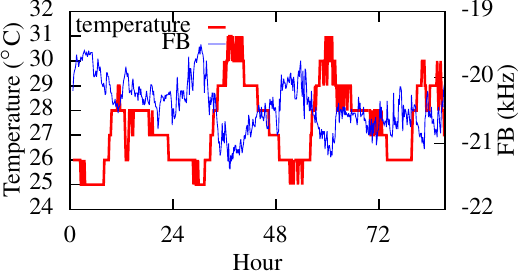}
    \label{fig:fb_temp}
  }
  \subfigure[CDF of FB variation.]
  {
    \includegraphics{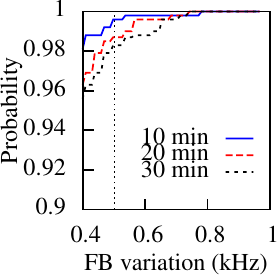}
    \label{fig:fb_cdf}
  }
  \caption{Temporal stability of FB over 87 hours.}
\end{figure}

FB can be affected by ambient condition such as temperature. We continuously track the FB of an SX1276-based end device for 87 hours to study its temporal stability. We place the end device with a temperature sensor in a semi-outdoor corridor with time-varying temperature.
The end device transmits 10 frames every 10 minutes to the $\mathsf{LoRaTS}$ gateway as shown in Fig.~\ref{fig:real_experiment_detailed}, resulting 1,440 frames per day. Fig.~\ref{fig:fb_temp} shows the end device's temperature and FB traces.
The Pearson correlation between FB and temperature is -0.78.
Moreover, the FB has transient variations that can be caused by interference from other communication systems operating in neighbor frequency bands. As $\mathsf{LoRaTS}$ detects the attack based on the changes of FB, such transient variations may cause false alarms. Fig.~\ref{fig:fb_cdf} shows the CDFs of the maximum FB variation if the end device transmits a frame every 10, 20, and 30 minutes. If the attack detection threshold is $500\,\text{Hz}$ based on our previous measurements of the additional FB introduced by the attack, from the CDFs, the false alarm rate (i.e., the probability that the FB variation exceeds $500\,\text{Hz}$) is about 0.4\%, 1.3\%, and 1.7\% for the three frame interval settings.
The SX1276 used in this paper does not have temperature compensated crystal oscillator (TCXO). For LoRa radios with TCXO, the false alarm rate can be further reduced.

\section{Discussions}
\label{sec:discuss}
{\bf Zero-FB attack:} To bypass the proposed attack detector, the adversary needs to precisely calibrate its eavesdropper and replayer to have FBs lower than the resolution of our FB estimation algorithm. Such calibration requires a highly accurate (e.g., ppb level) frequency source operating at the channel frequency, which is non-trivial. The GPSDO module of USRP provides a GPS-locked reference clock of $10\,\text{MHz}$ with 0.025 ppm accuracy \cite{gpsdo}. While the non-integer scaling from $10\,\text{MHz}$ to channel frequency may be subject to biases, the additional cost of two GPSDO modules (about US\$1,800) is non-trivial for the eavesdropper and replayer to tune frequency accurately.
There is also a possibility that the replayer's FB happens to cancel the eavesdropper's FB, rendering the superimposed FB zero. However, relying on such a random incident is an inefficient strategy for the attacker. Overall, the proposed low-cost (US\$25 for RTL-SDR) attack detector significantly increases the cost and technical barrier of attack.

{\bf Timestamp recovery:} Recovering timestamp under attack is challenging and needs further study. A recent concurrent LoRa demodulator \cite{ftrack} may not work for this purpose because it requires time-misalignment between two concurrent frames. The attacker can reduce the time-misalignment.


\section{Conclusion}
\label{sec:conclude}

This paper shows that sync-free data timestamping for LoRaWAN, though bandwidth-efficient, is susceptible to the easy-to-implement frame delay attack that can affect large areas.
To gain attack awareness, we design a gateway called $\mathsf{LoRaTS}$ that integrates a low-power SDR receiver with a commodity LoRaWAN gateway. The proposed least squares FB estimation algorithm achieves high resolution and can uncover the additional FBs introduced by the attack. In summary, with $\mathsf{LoRaTS}$, we can achieve efficient sync-free data timestamping with the awareness of the frame delay attack.


\section*{Acknowledgment}
 We acknowledge Zhenyu Yan for assistance in conducting the long-distance experiments in the NTU campus. We wish to thank our shepherd Dr. Julie McCann and the anonymous reviewers for providing valuable feedback on this work. This research was supported in part by two MOE AcRF Tier 1 grants (2019-T1-001-044 and 2018-T1-002-081).

\bibliographystyle{IEEEtran}
\balance
\bibliography{secure-ts}
\end{document}